\documentclass[aps,prl,reprint,superscriptaddress,floatfix,flushbottom,a4paper,nofootinbib,showpacs,amsmath]{revtex4-1}
\usepackage{amssymb}
\usepackage{dsfont}
\usepackage{graphicx}
\newcommand{\be}{\begin{equation}}
\newcommand{\ee}{\end{equation}}
\newcommand{\MS}{\overline{\mathrm{MS}}}
\newcommand{\MC}{\mathrm{MC}}

\newcommand{\G}{\mathrm{G}}
\newcommand{\latt}{\mathrm{latt}}
\newcommand{\nn}{\nonumber}
\newcommand{\lQ}{\Lambda_{\mathrm{QCD}}}
\newcommand{\al}{\alpha}
\begin{document}
\title
{Model-independent determination of the gluon condensate 
in four-dimensional SU(3) gauge theory} 
\author{Gunnar S.\ Bali}
\affiliation{Institut f\"ur Theoretische Physik, Universit\"at
Regensburg, D-93040 Regensburg, Germany}
\affiliation{Tata Institute of Fundamental Research, Homi Bhabha Road, Mumbai 400005, India}
\author{Clemens Bauer}
\affiliation{Institut f\"ur Theoretische Physik, Universit\"at
Regensburg, D-93040 Regensburg, Germany}
\author{Antonio Pineda}
\affiliation{Grup de F\'{\i}sica Te\`orica, Universitat
Aut\`onoma de Barcelona, E-08193 Bellaterra, Barcelona, Spain}
\date{\today}
\begin{abstract}
We determine the non-perturbative gluon condensate of four-dimensional 
SU(3) gauge theory in a model-independent way.
This is achieved by carefully subtracting high order perturbation theory
results from non-perturbative lattice QCD determinations
of the average plaquette. No indications of dimension two
condensates are found. The value of the gluon condensate turns out to be
of a similar size as the intrinsic ambiguity inherent to its
definition. We also determine the binding energy of a $B$ meson in the
heavy quark mass limit.
\end{abstract}
\maketitle

The operator product expansion (OPE)~\cite{Wilson:1969zs} is a fundamental
tool for theoretical analyses in quantum field theories.
Its validity is only proven
rigorously within perturbation theory, to arbitrary finite
orders~\cite{Zimmermann:1972tv}. The use of the OPE
in a non-perturbative framework was initiated
by the ITEP group~\cite{Vainshtein:1978wd}
(see also the discussion in~Ref.~\cite{Novikov:1984rf}), which postulated
that the OPE of a correlator could be approximated by the following series:
\be
\mathrm{correlator}(Q) \simeq \sum_d\frac{1}{Q^d}C_d(\alpha)
\langle O_d \rangle
\,,
\ee  
where the expectation values of local operators $O_d$
are suppressed by inverse powers of a large
external momentum $Q\gg\lQ$, according to their dimensionality $d$.
The Wilson coefficients $C_d(\alpha)$ 
encode the physics at momentum scales larger than $Q$.
These are well approximated by perturbative
expansions in the strong coupling parameter $\alpha$.
The large-distance
physics is described by the matrix elements
$\langle O_d \rangle$ that usually have to be
determined non-perturbatively.

Almost all QCD predictions of relevance
to particle physics phenomenology are based
on factorizations that are generalizations
of the above generic OPE.

For correlators where $O_0=\mathds{1}$, the first term of
the OPE expansion is a perturbative series in $\alpha$.
In pure gluodynamics the first
non-trivial gauge invariant local operator has dimension
four. Its expectation value is
the so-called non-perturbative gluon condensate
\begin{align}
\label{eq:GC}
\langle O_{\mathrm{G}}\rangle&=-\frac{2}{\beta_0}\left\langle\Omega\left| \frac{\beta(\alpha)}{\alpha}
G_{\mu\nu}^aG_{\mu\nu}^a\right|\Omega\right\rangle
\\
\nn
&=
\left\langle\Omega\left|  \left[1+\mathcal{O}(\alpha)\right]\frac{\alpha}{\pi}
G_{\mu\nu}^aG_{\mu\nu}^a\right|\Omega\right\rangle
\,.
\end{align}
This condensate plays a fundamental role in phenomenology, in particular
in sum rule analyses, as for many observables it is the
first non-perturbative OPE correction to the purely perturbative
result. In this Letter we will compute (and define) this object.
For this purpose we use the expectation value
of the plaquette calculated in Monte Carlo
(MC) simulations in lattice regularization with the
standard Wilson gauge action~\cite{Wilson:1974sk}
\begin{equation}
\langle P\rangle_{\mathrm{MC}}=\frac{1}{N^4}\sum_{x\in\Lambda_E}\langle P_x\rangle\,,
\end{equation}
where $\Lambda_E$ is a Euclidean spacetime lattice and
\begin{equation}
P_{x,\mu\nu}=1-\frac{1}{6}\mathrm{Tr}\left(U_{x,\mu\nu}+U_{x,\mu\nu}^{\dagger}\right)\,.
\end{equation}
For details on
the notation see Ref.~\cite{Bali:2014fea}.
The corresponding OPE reads
\be
\label{OPE}
\langle P \rangle_{\mathrm{MC}}=
\sum_{n=0}^{\infty}p_n\al^{n+1}
+\frac{\pi^2}{36}C_{\G}(\al)\,a^4\langle O_{\G}\rangle+\mathcal{O}\left(a^6\right)\,,
\ee
where $a$ denotes the lattice spacing.

The perturbative series is divergent due
to renormalons~\cite{Hooft} and other, subleading, instabilities.
This makes any determination of $\langle O_{\G}\rangle$ ambiguous,
unless we define how to truncate or how to
approximate the perturbative series. A reasonable definition
that is consistent with
$\langle O_{\G}\rangle \sim \lQ^4$ can only be given if the asymptotic
behaviour of the perturbative series is under control.
This has only been achieved recently~\cite{Bali:2014fea},
where the perturbative expansion of the plaquette was
computed up to $\mathcal{O}(\al^{35})$. The observed
asymptotic behaviour was in full compliance with renormalon
expectations, with successive contributions starting
to diverge for orders around $\al^{27}$--$\al^{30}$ within
the range of couplings $\al$ typically employed in present-day
lattice simulations. 

Extracting the gluon condensate from the
average plaquette was
pioneered in Refs.~\cite{Di Giacomo:1981wt,Kripfganz:1981ri,DiGiacomo:1981dp,Ilgenfritz:1982yx} and
many attempts followed during the next decades,
see, e.g., Refs.~\cite{Alles:1993dn,DiRenzo:1994sy,DiRenzo:1994sy,Ji:1995fe,DiRenzo:1995qc,Burgio:1997hc,Horsley:2001uy,Rakow:2005yn,Meurice:2006cr,Lee:2010hd,Horsley:2012ra}.
These suffered from insufficiently high perturbative orders and,
in some cases, also finite volume
effects. The failure to make contact to the asymptotic regime 
prevented a reliable lattice determination of $\langle O_{\G}\rangle$.
We solve this problem in this Letter. 

Truncating the infinite sum at the order
of the minimal contribution provides one definition of the perturbative series.
Varying the truncation order will result in changes of
size $\lQ^4a^4$, where the dimension $d=4$ is determined
by that of the gluon condensate. We
approximate the asymptotic series by the truncated sum
\be
\label{eq:truncate}
S_P(\al)\equiv S_{n_0}(\al)\,,\quad\mathrm{where}\quad S_n(\al)=\sum_{j=0}^{n}p_j\al^{j+1}\,.
\ee
$n_0\equiv n_0(\al)$ is the order for which $p_{n_0}\al^{{n_0}+1}$ is minimal.
We then obtain the gluon condensate from the relation
\begin{equation}
\langle O_{\G} \rangle=
\frac{36C_{\G}^{-1}(\al)}{\pi^2a^{4}(\al)}\left[\langle P \rangle_{\MC}(\al)-S_P(\al)\right]+
\mathcal{O}(a^2\lQ^2)
\,.
\label{eq:G2}
\end{equation} 

For the plaquette, the inverse Wilson coefficient
\begin{align}
C^{-1}_{\G}(\al)&=
\label{CP}
-\frac{2\pi\beta(\al)}{\beta_0\al^2}
\\
\nn
&=1+\frac{\beta_1}{\beta_0}\frac{\al}{4\pi}
+\frac{\beta_2}{\beta_0}\left(\frac{\al}{4\pi}\right)^2
+\frac{\beta_3}{\beta_0}\left(\frac{\al}{4\pi}\right)^3
+{\cal O}(\al^4)
\end{align}
is proportional to the
$\beta$-function~\cite{DiGiacomo:1990gy,DiGiacomo:1989id}.
For $j\leq 3$ the coefficients $\beta_j$  are known in the lattice
scheme (see Eq.~(25) of Ref.~\cite{Bali:2014fea}). The corrections to
$C_{\G}=1$ are small. However, the $\mathcal{O}(\al^2)$ and
$\mathcal{O}(\al^3)$ terms are of similar sizes. We will
account for this uncertainty in our error budget.

\begin{figure}
\includegraphics[width=.47\textwidth,clip=]{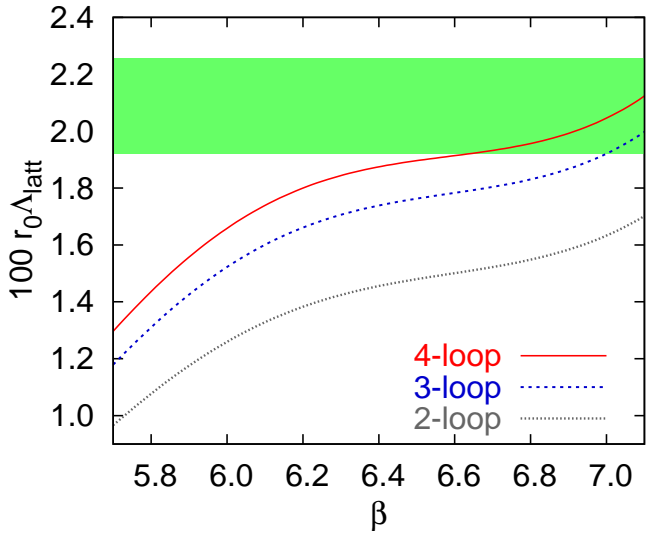}
\caption{Equation~(\protect\ref{eq:Necco}) over
Eq.~(\protect\ref{lambdapa}), truncated
at different orders. The
green band corresponds to
$r_0\Lambda_{\mathrm{latt}}=0.0209(17)$~\protect\cite{Capitani:1998mq}.
\label{fig:ratioLambda}}
\end{figure}
Integrating the $\beta$-function results in the
following dependence of the lattice spacing $a$ on
the coupling $\al$:
\begin{align}\label{lambdapa}
a&=
\frac{1}{\Lambda_{\latt}} \exp\left[-\frac{1}{t}
-b\ln\frac{t}{2}
+s_1 bt-s_2b^2t^2+\cdots\right]
\,,
\end{align}
where $t=\al\beta_0/(2\pi)$,
$b=\beta_1/(2\beta_0^2)$,
$s_1=(\beta_1^2-\beta_0\beta_2)/(4b\beta_0^4)$ and
$s_2=(\beta_1^3-2\beta_0\beta_1\beta_2+\beta_0^2\beta_3)/(16b^2\beta_0^6)$.
Equation~(\ref{lambdapa}) is not accurate in the lattice scheme
for typical $\beta$-values ($\beta\equiv 3/(2\pi\alpha)$) used in present-day simulations.
Instead, we employ the
phenomenological parametrization of Ref.~\cite{Necco:2001xg}
($x=\beta-6$)
\begin{align}
\label{eq:Necco}
a=r_0\exp\left(-1.6804-1.7331x+0.7849x^2-0.4428x^3\right)\,,
\end{align}
obtained by
interpolating non-perturbative lattice simulation results.
Equation~(\ref{eq:Necco}) was reported to be valid within an accuracy varying
from 0.5\% up to 1\% in the
range~\cite{Necco:2001xg} $5.7\leq\beta\leq 6.92$.
We plot the ratio of the above two equations $r_0\Lambda_{\latt}$ in 
Fig.~\ref{fig:ratioLambda}, where we truncate 
Eq.~(\ref{lambdapa}) at different orders. The green error band corresponds
to~\cite{Capitani:1998mq} $r_0 =0.0209(17)/\Lambda_{\latt}\simeq 0.5\,\mathrm{fm}$
($\Lambda_{\MS}\approx 28.809\Lambda_{\latt}$).
For large $\beta$-values this ratio should approach a constant.
Up to $\beta\approx 6.7$ this appears to be the case,
however, for $\beta>6.7$ the slope of the ratio starts to increase.
This may indicate violations of
Eq.~(\ref{eq:Necco})
for $\beta> 6.7$.
Therefore, we will restrict
ourselves to the range $\beta \in [5.8,6.65]$,
where $a(\beta)$ is given by Eq.~(\ref{eq:Necco}).
This corresponds to
$(a/r_0)^4 \in [3.1\times 10^{-5},5.5\times 10^{-3}]$, covering more than
two orders of magnitude.

\begin{figure}
\includegraphics[width=.48\textwidth,clip=]{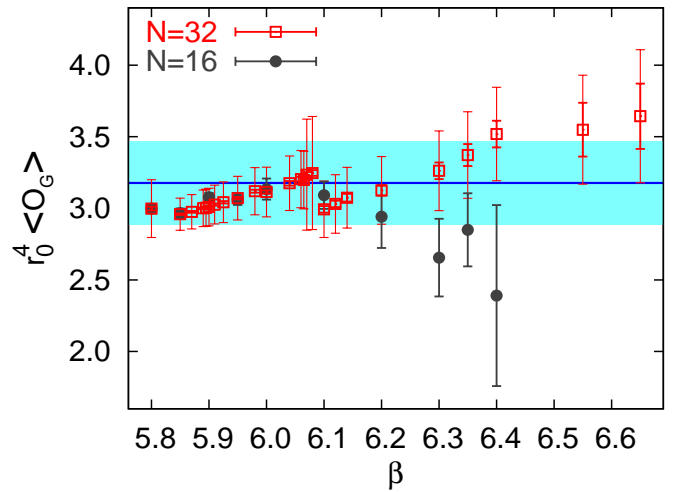}
\caption{Equation~(\protect\ref{eq:G2}) evaluated using the $N=16$ and $N=32$
MC data
of Ref.~\protect\cite{Boyd:1996bx}. The $N=32$ outer error bars include
the error of $S_P(\al)$. The error band is our prediction
for $\langle O_G \rangle$, Eq.~(\protect\ref{eq:G2final}).
\label{fig:PlaqMC}}
\end{figure}

Following Eq.~(\ref{eq:G2}), we subtract the
truncated sum $S_P(\al)$ calculated from the coefficients
$p_n$ of Ref.~\cite{Bali:2014fea} from the MC data on $\langle P \rangle_{\MC}(\al)$
of Ref.~\cite{Boyd:1996bx}. Multiplying this difference by
$36r_0^4/(\pi^2 C_{\G}a^4)$,
where $r_0/a$ is given by Eq.~(\ref{eq:Necco}),
gives $r_0^4\langle O_{\G}\rangle$
plus higher order non-perturbative terms. 
We show this combination in
Fig.~\ref{fig:PlaqMC}. The smaller error bars represent the
errors of the MC data, the outer error bars (not plotted
for $N=16$) the
total uncertainty, including that of $S_P$. This part of
the error is correlated between different $\beta$-values. 
The MC data
were obtained on volumes $N^4=16^4$ and $N^4=32^4$.
Towards large $\beta$-values the physical volumes
$N^4a^4(\beta)$
will become small, resulting in transitions
into the deconfined phase.
For $\beta<6.3$ we find no significant differences
between the $N=16$ and $N=32$ results. In the analysis
we restrict ourselves to the more precise
$N=32$ data and, to keep finite size effects under control,
to $\beta\leq 6.65$. We also 
limit ourselves to $\beta\geq 5.8$ to avoid large
$\mathcal{O}(a^2)$ corrections. 
At very large $\beta$-values not only does the parametrization Eq.~(\ref{eq:Necco})
break down but obtaining meaningful results becomes challenging
numerically: the individual errors both of $\langle P\rangle_{\MC}(\al)$
and of $S_P(\al)$ somewhat decrease with increasing $\beta$.
However, there are strong cancellations between these two terms, in
particular at large $\beta$-values, since this difference
decreases with $a^{-4}
\sim\Lambda_{\latt}^4\exp(16\pi^2\beta/33)$ on dimensional
grounds while $\langle P\rangle_{\MC}$ depends only logarithmically
on $a$.

The coefficients $p_n$ of $S_P(\al)$
were obtained in Ref.~\cite{Bali:2014fea}. The $p_n$ carry statistical errors
and successive orders are correlated. Using the covariance matrix, also
obtained in Ref.~\cite{Bali:2014fea}, the statistical error of $S_P(\al)$ can
be calculated. In that reference, coefficients $p_n(N)$ were first
computed on finite volumes of $N^4$ sites and subsequently extrapolated
to their infinite volume limits $p_n$. This extrapolation
is subject to parametric uncertainties that need to be
estimated. We follow Ref.~\cite{Bali:2014fea}
and add the differences between determinations
using $N\geq \nu$ points for $\nu=9$ (the central values)
and $\nu=7$ as systematic errors to our statistical errors.

\begin{figure}
\includegraphics[width=.48\textwidth,clip=]{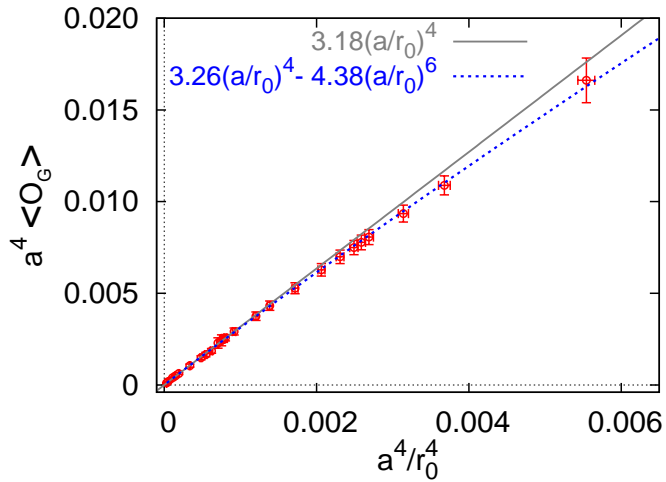}
\caption{Equation~(\protect\ref{eq:G2}) times $a^4$ vs.\ $a^4(\al)/r_0^4$
from Eq.~(\protect\ref{eq:Necco}). The linear fit
with slope Eq.~(\protect\ref{eq:G2final}) is to $a^4< 0.0013\,r_0^4$
points only.
\label{linear}}
\end{figure}

The data in Fig.~\ref{fig:PlaqMC} show an approximately
constant behaviour.\footnote{Note that
$n_0$ increases from 26 to 27 at $\beta=5.85$, from
27 to 28 at $\beta=6.1$ and from 28 to 29
at $\beta=6.55$. This quantization of $n_0$ explains the visible jump
at $\beta=6.1$.} This indicates
that, after subtracting
$S_P(\al)$ from the corresponding MC values
$\langle P \rangle_{\MC}(\al)$, the remainder scales like $a^4$.
This can be seen more explicitly in Fig.~\ref{linear}, where
we plot this difference in lattice units against $a^4$.
The result is consistent with a linear behaviour
but a small curvature seems to be present that can be parametrized as
an $a^6$-correction.
The right-most point ($\beta=5.8$) corresponds to $a^{-1}\simeq
1.45$~GeV while $\beta=6.65$ corresponds to
$a^{-1}\simeq 5.3$~GeV. Note that
$a^2$-terms are clearly ruled out.  

We now determine the gluon condensate.
We obtain the central value and its statistical error
$\langle O_{\G}\rangle=3.177(36)r_0^{-1}$
from averaging the $N=32$ data for $6.0\leq\beta\leq 6.65$.
We now estimate the systematic uncertainties.
Different infinite volume extrapolations of
the $p_n(N)$ data~\cite{Bali:2014fea}
result in changes of the prediction of about $6\%$.
Another $6\%$ error is due to including an $a^6$-term or not
and varying the fit range.
Next there is a scale error of
about 2.5\%, translating $a^4$ into units of $r_0$.
The uncertainty of the perturbatively determined
Wilson coefficient $C_{\G}$ is of a similar size.
This is estimated as the difference between
evaluating Eq.~(\ref{CP}) to $\mathcal{O}(\al^2)$
and to $\mathcal{O}(\al^3)$.
Adding all these sources of uncertainty in quadrature
and using~\cite{Capitani:1998mq} $\Lambda_{\MS}=0.602(48)r_0^{-1}$
yields
\be
\label{eq:G2final}
\langle O_{\G} \rangle=3.18(29)r_0^{-4}=24.2(8.0)\Lambda_{\MS}^4\,.
\ee 

The gluon condensate of Eq.~(\ref{eq:GC}) is
independent of the renormalization scale.
However, $\langle O_{\G}\rangle$ was obtained employing one
particular prescription in terms of the observable
and our choice of how to truncate
the perturbative series within a given renormalization scheme.
Different (reasonable) prescriptions can in principle
give different results. One may for instance choose to truncate 
the sum at orders $n_0(\al) \pm \sqrt{n_0(\al)}$ and
the result would still scale like $\lQ^4$. 
We estimated this intrinsic ambiguity of the definition of the
gluon condensate in Ref.~\cite{Bali:2014fea}
as $\delta\langle O_{\G}\rangle=
36/(\pi^2C_{\G}a^4)\,\sqrt{n_0}p_{n_0}\alpha^{n_0+1}$, i.e.\
as $\sqrt{n_0(\al)}$
times the contribution of the minimal term,
\be
\label{eq:prescript}
\delta\langle O_{\G}\rangle=27(11)\Lambda_{\MS}^4\,.
\ee
Up to $1/n_0$-corrections this definition is scheme- and
scale-independent and corresponds
to the (ambiguous) imaginary part of the Borel integral times
$\sqrt{2/\pi}$.

In QCD with sea quarks
the OPE of the average plaquette or of the Adler function
will receive additional contributions from the chiral condensate.
For instance $\langle O_{\G}\rangle$ needs to be redefined,
adding terms
$\propto\langle \gamma_m(\al)m\bar\psi\psi\rangle$~\cite{Tarrach:1981bi}.
Due to this and the problem of setting a physical scale
in pure gluodynamics, it is difficult to assess the precise
numerical impact
of including sea quarks onto our estimates
\be
\label{eq:G2GeV}
\langle O_{\G} \rangle\simeq 0.077\, \mathrm{GeV}^4\,,
\quad
\delta\langle O_{\G}\rangle\simeq 0.087\, \mathrm{GeV}^4\,,
\ee 
which we obtain using $r_0\simeq 0.5$~fm~\cite{Sommer:1993ce}.
While the systematics of applying
Eqs.~(\ref{eq:G2final})--(\ref{eq:prescript})
to full QCD are unknown, our main observations
should still extend to this case.
We remark that our prediction of the gluon condensate of
Eq.~(\ref{eq:G2GeV}) is significantly bigger than values
 obtained in one- and two-loop sum rule analyses,
ranging from
0.01~GeV${}^4$~\cite{Vainshtein:1978wd,Ioffe:2002be} up
to 0.02~GeV${}^4$~\cite{Broadhurst:1994qj,Narison:2011xe}.
However, these numbers were not extracted in the
asymptotic regime, which for a $d=4$ renormalon we expect to set in
at orders $n \gtrsim 7$ for the $\MS$ scheme. 
Moreover, we remark that in schemes
without a hard ultraviolet cut-off, like dimensional regularization,
the extraction of $\langle O_{\G}\rangle$ can become obscured by the
possibility of ultraviolet
renormalons. Independent of these considerations,
all these values are smaller than the intrinsic
prescription dependence Eq.~(\ref{eq:prescript}).

\begin{figure}
\includegraphics[width=.47\textwidth,clip=]{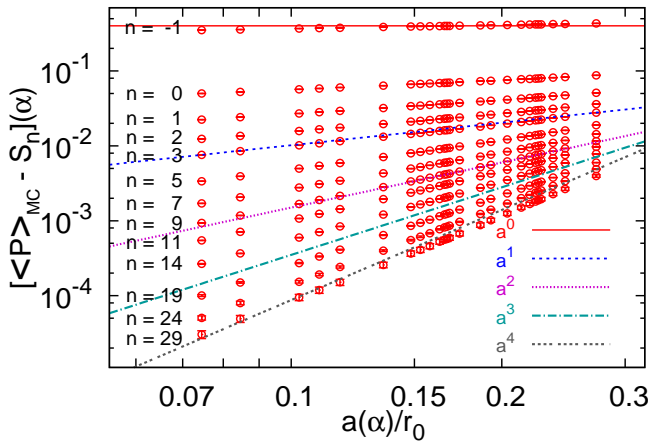}
\caption{Differences $\langle P\rangle_{\MC}(\al)-S_n(\al)$
between MC data and sums truncated
at orders $\al^{n+1}$ ($S_{-1}=0$)
vs.\ $a(\al)/r_0$. The lines $\propto a^j$ are drawn
to guide the eye.
\label{fig:a2}}
\end{figure}

Our analysis confirms the validity of the OPE beyond perturbation theory 
for the case of the plaquette. Our $a^4$-scaling clearly
disfavours suggestions about the existence of
dimension two condensates beyond the standard OPE
framework~\cite{Burgio:1997hc,Chetyrkin:1998yr,Gubarev:2000nz,RuizArriola:2006gq,Andreev:2006vy}.
In fact we can also explain why
an $a^2$-contribution to the plaquette was found in Ref.~\cite{Burgio:1997hc}.
In the log-log plot of Fig.~\ref{fig:a2} we subtract sums $S_n$, truncated at
different fixed orders $\al^{n+1}$, from $\langle P\rangle_{\MC}$.
The scaling continuously
turns from $\sim a^0$ at $\mathcal{O}(\al^0)$ to $\sim a^4$ around
$\mathcal{O}(\al^{30})$. Note that truncating at an
$\al$-independent fixed order is inconsistent, explaining why we
never exactly obtain an $a^4$-slope.
For $n\sim 9$ we reproduce the $a^2$-scaling reported
in Ref.~\cite{Burgio:1997hc} for a fixed order truncation
at $n=7$.
In view of Fig.~\ref{fig:a2} we conclude that
the observation of this scaling power was accidental.

\begin{figure}
\includegraphics[width=.47\textwidth,clip=]{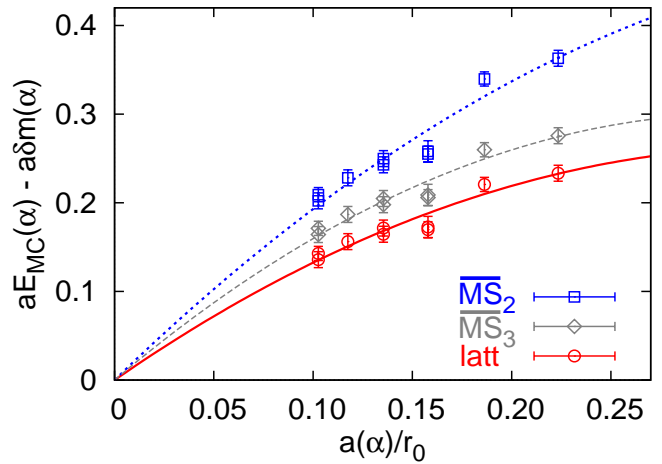}
\caption{$aE_{\MC}-a\delta m$ vs. $a/r_0$. The expansion
of $a\delta m$ was also converted into the $\MS$ scheme at
two ($\MS_2$) and three ($\MS_3$) loops. The curves are fits to
$\overline{\Lambda} a+c a^2$.
\label{fig:lambdabar}}
\end{figure}

The methods used in this Letter can be applied to other observables.
As an example we analyse the binding energy
$\overline{\Lambda}=E_{\MC}(\al)-\delta m(\al)$~\cite{Luke:1990eg,Falk:1992fm,Crisafulli:1995pg}
of heavy quark effective theory.
The perturbative expansion of $a\delta m(\al)=\sum_n c_n\al^{n+1}$
was obtained in Refs.~\cite{Bauer:2011ws,Bali:2013pla}
up to $\mathcal{O}(\al^{20})$, and its intrinsic ambiguity
$\delta\overline{\Lambda}=\sqrt{n_0}c_{n_0}\alpha^{n_0+1}=0.748(42)\Lambda_{\MS}=0.450(44)r_0^{-1}$ in Refs.~\cite{Bali:2013pla,Bali:2013qla}.
MC data for the ground state energy $E_{\MC}$ of a static-light meson
with the Wilson gauge action can be found
in Refs.~\cite{Duncan:1994uq,Allton:1994tt,Ewing:1995ih}. While for the
gluon condensate we expected an $a^4$-scaling (see Fig.~\ref{linear}),  
for $aE_{\MC}(\al)-a\delta m(\al)$ we expect a scaling linear in $a$.
Comforting enough this is what we find, up
to $a\mathcal{O}(a)$ discretization corrections,
see Fig.~\ref{fig:lambdabar}. 
Subtracting the partial sum truncated at orders $n_0(\al)= 6$
from the $\beta\in[5.9,6.4]$
data, we obtain $\overline{\Lambda}= 1.55(8) r_0^{-1}$
from such a linear plus quadratic fit, where
we only give the statistical uncertainty.
The errors of the perturbative coefficients are all
tiny, which allows us to transform
the expansion $a\delta m(\al)$
into $\MS$-like schemes and to compute $\overline{\Lambda}$ accordingly. 
We define the schemes $\MS_2$ and $\MS_3$ by truncating
$\al_{\MS}(a^{-1})=\al(1+d_1\al+d_2\al^2+\ldots)$ exactly at
$\mathcal{O}(\al^3)$ and 
$\mathcal{O}(\al^4)$, respectively. The $d_j$
are known for $j\leq 3$~\cite{Bali:2013pla,Bali:2013qla}.
We typically find $n_0^{\MS_i}(\al_{\MS_i})=2, 3$
and obtain $\overline{\Lambda} \sim 2.17(8) r_0^{-1}$
and $\overline{\Lambda} \sim 1.89(8) r_0^{-1}$,
respectively, see Fig.~\ref{fig:lambdabar}.
We conclude that the changes due to these resummations are
indeed of the size
$\delta\overline{\Lambda}\sim 0.5 r_0^{-1}$,
adding confidence that our definition of the
ambiguity is neither a gross overestimate nor an underestimate.
For the plaquette, where we expect $n_0^{\MS} \sim 7$,
we cannot carry out a similar analysis,
due to the extremely high precision that is required
to resolve the differences 
between $S_P(\al)$ and $\langle P \rangle_{\MC}(\al)$, which
largely cancel in Eq.~(\ref{eq:G2}).

In conclusion, for the first time ever, perturbative expansions at orders
where the asymptotic regime is reached have been subtracted
from non-perturbative MC data of the static-light meson mass
and of the plaquette, thereby validating the OPE beyond perturbation
theory. The scaling of the latter difference with the lattice spacing
confirms the dimension $d=4$. Dimension $d<4$ slopes appear only
when subtracting the perturbative series truncated at fixed
pre-asymptotic orders: lower dimensional ``condensates'' discussed in the
literature, see, e.g., Refs.~\cite{Chetyrkin:1998yr,Gubarev:2000nz,RuizArriola:2006gq,Andreev:2006vy}, are just approximate parametrizations of
unaccounted perturbative effects, i.e.\ of the short-distance behaviour,
and, thus, observable-dependent
(unlike the non-perturbative gluon condensate). Such
simplified parametrizations
introduce unquantifiable errors and, therefore, are of limited
phenomenological use.

We have obtained an accurate value of
the gluon condensate in SU(3) gluodynamics, Eq.~(\ref{eq:G2final}).
It is of a similar size as the intrinsic
difference, Eq.~(\ref{eq:prescript}), between
(reasonable) subtraction prescriptions. This result contradicts the 
implicit assumption of sum rules analyses that the renormalon ambiguity is 
much smaller than leading non-perturbative corrections.
The value of the gluon condensate obtained with sum rules
can vary significantly due to this intrinsic, renormalization
scheme-independent, ambiguity if determined using different
prescriptions or truncating at different orders in perturbation theory.
Clearly, the impact of this, e.g., on determinations
of $\alpha_s$ from $\tau$-decays or from lattice simulations needs to be
assessed carefully.

Finally, the inherent ambiguity  of (reasonable)
definitions of the static-light meson mass was estimated
in Refs.~\cite{Bali:2013pla,Bali:2013qla}.
Here, in a combined analysis, this estimate was confronted with
MC data and confirmed.

\begin{acknowledgments}
We thank Taekoon Lee for finding a typo in an earlier
version of the manuscript.
This work was supported by the German DFG
Grant No.~SFB/TRR-55, the Spanish 
Grants No.~FPA2010-16963 and No.~FPA2011-25948, the Catalan Grant
No.~SGR2009-00894
and EU ITN STRONGnet 238353.
\end{acknowledgments}

\end{document}